\documentclass[amsmath,amssymb,12pt,superscriptaddress,nofootinbib]{revtex4}

\usepackage{graphicx}
\usepackage{bm}

\newcommand{\delr}{\partial_r}
\newcommand{\delt}{\partial_t}
\newcommand{\tb}{t_{\rm B}}

\begin{document}
\thispagestyle{empty}

{\baselineskip0pt
\leftline{\baselineskip14pt\sl\vbox to0pt{
               \hbox{\it Yukawa Institute for Theoretical Physics} 
              \hbox{\it Kyoto University}
               \vspace{1mm}
              \hbox{\it Department of Mathematics and Physics}
               \hbox{\it Osaka City  University}
               \vss}}
\rightline{\baselineskip16pt\rm\vbox to20pt{
            \hbox{YITP-10-87}
            {\hbox{OCU-PHYS-340}
            \hbox{AP-GR-84}
            }
\vss}}%
}

\author{Chul-Moon Yoo}\email{yoo@yukawa.kyoto-u.ac.jp}
\affiliation{
Yukawa Institute for Theoretical Physics, Kyoto University
Kyoto 606-8502, Japan
}
\author{Tomohiro Kai}\email{hirotomo203@gmail.com}
\affiliation{ 
Department of Mathematics and Physics,
Graduate School of Science, Osaka City University,
3-3-138 Sugimoto, Sumiyoshi, Osaka 558-8585, Japan
}
\author{Ken-ichi Nakao}\email{knakao@sci.osaka-cu.ac.jp}
\affiliation{ 
Department of Mathematics and Physics,
Graduate School of Science, Osaka City University,
3-3-138 Sugimoto, Sumiyoshi, Osaka 558-8585, Japan
}

\vskip2cm
\title{Redshift Drift in LTB Void Universes}

\begin{abstract}
We study the redshift drift, i.e., the time derivative of the cosmological redshift 
in the Lema\^itre-Tolman-Bondi (LTB) solution  
in which the observer is assumed to be located at the symmetry center. 
This solution has often been studied 
as an anti-Copernican 
universe model to explain the acceleration of 
cosmic volume expansion without introducing the concept of dark energy. 
One of decisive differences between LTB universe models and Copernican 
universe models with dark energy is believed to be 
the redshift drift. The redshift drift is negative in all known LTB universe models, 
whereas it is positive in the redshift domain $z \lesssim 2$  in
Copernican models with dark energy. 
However, there have been no detailed studies on this subject. 
In the present paper, we prove that the redshift drift of an off-center source is always negative  
in the case of LTB void models. 
We also show 
that the redshift drift can be positive with 
an extremely large hump-type inhomogeneity. 
Our results suggest that we can determine whether 
we live near the center of a large void 
without dark energy by observing the redshift drift. 

\end{abstract}

\maketitle

\section{introduction}

The standard cosmological model is based on the so-called Copernican principle 
that we are not located in a special position in the universe. 
This model can naturally explain almost all observational data, and  
consequently seems to imply that the Copernican principle is a reality. 
However, we should not blindly rely on this principle 
without observational justifications. 

The highly isotropic cosmic microwave background (CMB) implies the 
isotropy of our universe at our position. 
Then, if we assume that
the Copernican principle is correct, we necessarily arrive at the conclusion that 
our universe is isotropic at every position, or equivalently, 
our universe is homogeneous and isotropic. 
By virtue of the homogeneity and isotropy, the standard cosmological model 
is determined by several parameters called the cosmological parameters.  
In order to determine the cosmological parameters, 
observational data are interpreted under the assumption 
of a homogeneous and isotropic universe on average. 
Here, we should note that it is not clear at all how large the 
systematic errors would be in the determination of the cosmological parameters, 
if the Copernican principle is abandoned.
Thus, it is an unavoidable task in observational cosmology to investigate possible 
``anti-Copernican'' universe  models and test if such models can
be observationally excluded. 

Almost all anti-Copernican universe models are based on the 
Lema\^itre-Tolman-Bondi (LTB) solution which describes the dynamics of 
a spherically symmetric dust.  
In anti-Copernican universe models, 
an observer like us is usually assumed to be located 
in the neighborhood of the symmetry center~\cite{Zehavi:1998gz,
Tomita:1999qn,Tomita:2000jj,Tomita:2001gh,Celerier:1999hp}. 
In recent years, such models have attracted much 
attention, since some LTB universe models 
can recover the observed 
distance-redshift relation without introducing 
dark energy~\cite{Alnes:2005rw,Alnes:2006pf,Alnes:2006uk,
Alexander:2007xx,Bolejko:2008cm,Enqvist:2007vb,Enqvist:2006cg,
February:2009pv,GarciaBellido:2008nz,GarciaBellido:2008yq,
GarciaBellido:2008gd,Garfinkle:2006sb,Kasai:2007fn,
2000ApJ...529...26T,Tomita:2000rf,Zibin:2008vk,Clifton:2009kx}, 
and various ways to observationally test these models
have been proposed by many authors~\cite{Biswas:2007gi,
Bolejko:2005fp,Bolejko:2008xh,Bolejko:2008ya,
Brouzakis:2006dj,Brouzakis:2007zi,Bhattacharya:2009bz,Clarkson:2007pz,
Caldwell:2007yu,Clifton:2008hv,Dabrowski:1997sm,Goodman:1995dt,
Godlowski:2004gh,Jia:2008ti,Lasky:2010vn,Marra:2007pm,Moffat:2006ct,
PascualSanchez:1999zr,Quartin:2009xr,Romano:2007zz,Stelmach:2006zc,
Tanimoto:2009mz,Tomita:2009yx,Uzan:2008qp,
Yoo:2010qy,Clarkson:1999yj,Zibin:2008vj,Regis:2010iq,Clarkson:2010ej,Clarkson:2010uz,
Biswas:2010xm,Kodama:2010gr,Foreman:2010uj,Moss:2010jx,Marra:2010pg,Zhang:2010fa}.

As shown by previous studies, 
the LTB solution can explain various observational data besides the 
distance-redshift relation without the need to introduce dark energy. 
This is because this solution 
has functional degrees of freedom with respect to the comoving radial 
coordinate. 
In order to check the LTB universe models observationally, 
it is crucial to find observable quantities which can reveal  
differences between the LTB universe models and Copernican 
universe models with the dark energy. 
One such quantity 
is believed to be the redshift drift, 
i.e., the time derivative of the cosmological 
redshift~\cite{Uzan:2008qp}. 
In the case of the $\Lambda$CDM model, 
which is the most likely 
Copernican model at present, the redshift drift is positive in the 
redshift domain $z\lesssim2$, since the cosmological constant $\Lambda$ 
causes repulsive gravity. 
By contrast, there is no exotic matter 
with the violation of the strong energy condition in the LTB solution. 
Thus, as long as there is no highly inhomogeneous structure, 
the redshift drift might be negative in LTB universe models. 
Although several authors have pointed out the importance of the 
redshift drift~\cite{Uzan:2008qp,Dunsby:2010ts,Quartin:2009xr,Yoo:2008su}, 
there has been no detailed study of its general behavior  
in LTB universe models. 
It is the purpose of this paper to investigate it. 

This paper is organized as follows. 
In Sec.~\ref{sec:ltb}, 
we briefly review the LTB solution. 
In Sec.~\ref{sec:basicEq}, we derive the equation 
for the redshift drift and 
show the behaviour of the redshift drift 
near an observer located at the 
symmetry center in LTB universe models. 
In Sec.~\ref{sec:negative}, we define LTB void models and 
prove a theorem on the redshift drift in these models. 
In Sec.~\ref{sec:positive}, we show that the redshift drift can be 
positive even in an LTB universe model, 
if an extremely large 
hump-type mass density distribution 
exists. 
Sec.~\ref{sec:conclusions} 
is devoted to the summary and discussion. 

In this paper, we denote the speed of light 
and Newton's gravitational constant by $c$ and $G$, respectively. 

\section{the LTB solution}
\label{sec:ltb}
As mentioned in the introduction,
we consider a spherically symmetric 
inhomogeneous universe filled with dust. 
This universe is described by an exact solution of 
the Einstein equations, which is known as the 
Lema\^itre-Tolman-Bondi (LTB) solution. 
The metric of the LTB solution is given by
\begin{equation}
ds^2=-c^2dt^2+\frac{\left(\partial_r R(t,r)\right)^2}{1-k(r)r^2}dr^2
+R^2(t,r)d\Omega^2, \label{eq:metric}
\end{equation}
where $k(r)$ is an arbitrary function of the radial coordinate $r$. 
The matter is dust whose stress-energy tensor is given by
\begin{equation}
T^{\mu\nu}=\rho u^\mu u^\nu,
\end{equation}
where $\rho=\rho(t,r)$ is the mass density, 
and $u^\mu$ is the four-velocity of 
the fluid element. 
The coordinate system in Eq.~(\ref{eq:metric}) is chosen
in such a way that $u^\mu=(c,0,0,0)$.

The circumferential radius $R(t,r)$ 
is determined by one of the Einstein equations, 
\begin{equation}
\left(\frac{\partial R}{\partial t}\right)^2=\frac{2GM(r)}{R}-c^2kr^2,
\label{eq:Einstein-eq}
\end{equation}
where $M(r)$ is an arbitrary function related to 
the mass density $\rho$ by
\begin{equation}
\rho(t,r)=\frac{1}{4\pi R^2\delr R}\frac{dM}{dr}.
\label{eq:rho}
\end{equation}
$M(r)$ is known as the Misner-Sharp mass that is the quasi-local mass  
naturally introduced into the spherically symmetric spacetime\cite{Misner:1964je}.
Here, it should be noted that 
the Misner-Sharp mass is not necessarily a non-decreasing function of 
the comoving radial coordinate $r$, even if 
the mass density $\rho$ is non-negative. 
In the case that a hypersurface labeled by $t$ is a homogeneous and isotropic 
space with positive curvature, $r$ can be chosen so that the 
circumferential radius is given by $R=a(t)\sin r$, where $a(t)$ is a positive function 
of time (in this case, $k(r)=r^{-2}\sin^2r$). Thus, $\delr R=a(t)\cos r$ is positive 
for $0\leq r<\pi/2$, whereas it is negative $\pi/2<r\leq \pi$. 
Since $\rho$ is 
spatially constant by assumption, 
the derivative of the Misner-Sharp mass $dM/dr$ 
is positive for 
$0\leq r<\pi/2$, whereas it is negative for $\pi/2<r\leq \pi$.
However, in this paper, we assume that the Misner-Sharp mass is a monotonically 
increasing function of $r$ in the domain of interest. 
This assumption is equivalent to the one that
 $\delr R$ is positive if $\rho$ is positive.

Following Ref.~\cite{Tanimoto:2007dq}, we write the solution 
of Eq.~(\ref{eq:Einstein-eq}) in the form,
\begin{eqnarray}
R(t,r)&=&(6GM)^{1/3}[t-t_{\rm B}(r)]^{2/3} S(x), 
\label{eq:YS}\\
x&=&c^2kr^2\left(\frac{t-t_{\rm B}}{6GM}\right)^{2/3}, \label{eq:defx}
\end{eqnarray}
where $t_{\rm B}(r)$ is an arbitrary function 
which determines the big bang time, 
and $S(x)$ is a function defined implicitly as
\begin{equation}
S(x)=
\left\{\begin{array}{lll}
\displaystyle
\frac{\cosh\sqrt{-\eta}-1}{6^{1/3}(\sinh\sqrt{-\eta}
-\sqrt{-\eta})^{2/3}}
\,;\qquad
&\displaystyle
x=\frac{-(\sinh\sqrt{-\eta}-\sqrt{-\eta})^{2/3}}{6^{2/3}}
\quad&\mbox{for}~~x\leq0\,,
\\
\displaystyle
\frac{1-\cos\sqrt{\eta}}{6^{1/3}(\sqrt{\eta}
-\sin\sqrt{\eta})^{2/3}}
\,;&\displaystyle
x=\frac{(\sqrt{\eta}-\sin\sqrt{\eta})^{2/3}}{6^{2/3}}
\quad&\mbox{for}~~x>0\,.
\end{array}\right.
\label{eq:defS}
\end{equation}
The function $S(x)$ is analytic for $x<(\pi/3)^{2/3}$. 
Some characteristics of the function $S(x)$ 
are given in Refs.~\cite{Yoo:2008su} and \cite{Tanimoto:2007dq}. 

As shown in the above, the LTB solution has three arbitrary functions, 
$k(r)$, $M(r)$ and $t_{\rm B}(r)$. 
One of them is a gauge degree of freedom 
for the rescaling of $r$.
In this paper, since $M$ is assumed to be a monotonically increasing function of $r$, 
we can fix this freedom by setting
\begin{equation}
M=\frac{4}{3}\pi\rho_0r^3, \label{eq:mass}
\end{equation}
where $\rho_0$ is the mass density at the 
symmetry center at the present time $t_0$, i.e., $\rho_0=\rho(t_0,0)$.
By this choice, we have
\begin{equation}
R(t_0,r)=r+{\cal O}(r^2). \label{eq:R}
\end{equation}
As in the case of the homogeneous and isotropic 
universe, the present Hubble parameter $H_0$ is related to $\rho_0$ as 
\begin{equation}
H_0^2+k(0)c^2=\frac{8}{3}\pi G\rho_0. 
\end{equation}

\section{Equation for the redshift drift}
\label{sec:basicEq}

In order to study the cosmological redshift and the redshift drift, 
we consider ingoing radial null geodesics. 
The cosmological redshift $z$ of a light ray from 
a comoving source at $r$ to the observer at the symmetry 
center  $r=0$ is defined by
\begin{equation}
z(r):=\frac{k^t\left(\lambda\left(r\right)\right)}{k^t\left(\lambda\left(0\right)\right)}-1,
\end{equation}
where $k^t$ is the time component of the null geodesic tangent, and 
$\lambda$ is the affine parameter which can be regarded as a function of $r$. 
From the geodesic equations, we have the equation for the redshift $z$ as 
\begin{equation}
\frac{dz}{dr}=\frac{(1+z)\delt\delr R}{c\sqrt{1-kr^2}}.
\label{eq:dzdr}
\end{equation}
The null condition leads to
\begin{equation}
\frac{dt}{dr}=-\frac{\delr R}{c\sqrt{1-kr^2}}.
\label{eq:dtdr}
\end{equation}

We denote the trajectories of light rays observed by the central observer at 
$t=t_0$ and $t=t_0+\delta t_0$, respectively, by 
\begin{equation}
\left\{
\begin{array}{c}
z=z_{\rm lc}(r;t_0)\\
t=t_{\rm lc}(r;t_0)
\end{array}
\right.
\label{eq:geo1}
\end{equation}
and
\begin{equation}
\left\{
\begin{array}{c}
z=z_{\rm lc}(r;t_0+\delta t_0)=:z_{\rm lc}(r;t_0)+\delta z(r)\\
t=t_{\rm lc}(r;t_0+\delta t_0)=:t_{\rm lc}(r;t_0)+\delta t(r)
\end{array}
\right..
\label{eq:geo2}
\end{equation}
Here, by their definitions, we have $t_{\rm lc}(0;t_0)=t_0$, 
$z_{\rm lc}(0;t)=0$, $\delta z(0)=0$ and $\delta t(0)=\delta t_0$. 
Substituting Eq.~\eqref{eq:geo2} into Eqs.~\eqref{eq:dzdr} 
and \eqref{eq:dtdr}, and regarding $\delta z(r)$ and 
$\delta t(r)$ as infinitesimal quantities, we obtain  
\begin{eqnarray}
\frac{d}{dr}\delta z&=&\frac{\delt\delr R}{c\sqrt{1-kr^2}}\delta z
+\frac{(1+z)\delt^2\delr R}{c\sqrt{1-kr^2}}\delta t, \\
\frac{d}{dr}\delta t&=&\frac{-\delt\delr R}{c\sqrt{1-kr^2}}\delta t, 
\end{eqnarray}
where we have used the fact that 
\eqref{eq:geo1} satisfies Eqs.~\eqref{eq:dzdr} 
and \eqref{eq:dtdr}, and 
the arguments of $\delt\delr R$ and $\delt^2 \delr R$ are 
$t=t_{\rm lc}(r,t_0)$ and $r$. 

Hereafter, we consider the case where the cosmological redshift $z$ is 
monotonically increasing with $r$. 
We say that such a model is {\it $z$-normal}. 
Then, we replace the independent variable $r$ by $z=z_{\rm lc}(r;t_0)$. 
By using 
\begin{equation}
\frac{d}{dr}=\frac{dz}{dr}\frac{d}{dz}
=\frac{(1+z)\delr\delr R}{c\sqrt{1-kr^2}}\frac{d}{dz},
\end{equation}
we have
\begin{eqnarray}
\frac{d}{dz}\delta z&=&\frac{\delta z}{1+z}
+\frac{\delt^2\delr R}{\delt \delr R}\delta t, 
\label{eq:delz}
\\
\frac{d}{dz}\delta t&=&-\frac{\delta t}{1+z}. 
\label{eq:delt}
\end{eqnarray}
We can easily integrate Eq.~\eqref{eq:delt} to obtain 
\begin{equation}
\delta t=\frac{\delta t_0}{1+z}.
\end{equation}
By using the above result, Eq.~\eqref{eq:delz} is rewritten 
in the following form
\begin{equation}
\frac{d}{dz}\left(\frac{\delta z}{1+z}\right)
=\frac{1}{(1+z)^2}\frac{\delt^2\delr R}{\delt \delr R}
\delta t_0.
\label{eq:delzfin}
\end{equation}

Here, let us study the redshift drift $\delta z$ 
in the neighborhood of the symmetry center. 
By the regularity at $r=0$, we have
\begin{eqnarray}
k(r)&=&k_0+\mathcal O(r), \\
\tb(r)&=&\mathcal O(r), 
\end{eqnarray}
where, by using the freedom of the constant time translation, we 
have set $\tb(0)=0$. 
From Eq.~\eqref{eq:Einstein-eq}, we have
\begin{eqnarray}
\left.\left(\frac{\delt R}{R}\right)^2\right|_{t=t_0,r=0}
&=&-c^2k_0+\frac{8\pi G\rho_0}{3}=H_0^2, \\
\left.\delt \delr R\right|_{t=t_0,r=0}&=&H_0, \\
\left.\delt^2\delr R\right|_{t=t_0,r=0}&=&-\frac{4\pi G\rho_0}{3}
=-\frac{1}{2}\Omega_{\rm m0}H_0^2, 
\end{eqnarray}
where we have used Eq.~\eqref{eq:R} and $\Omega_{\rm m0}=8\pi G\rho/3H_0^2$. 
In the neighborhood of $r=0$, we see from Eq.\eqref{eq:delzfin} that 
\begin{equation}
\left.\frac{d}{dz}\delta z\right|_{t=t_0,r=0}
=\left.\frac{\delt^2\delr R}{\delt\delr R}\right|_{t=t_0,r=0}\delta t_0
+\mathcal O(z)
=-\frac{1}{2}\Omega_{\rm m0}\delta t_0+\mathcal O(z). 
\end{equation}
Thus, we have
\begin{equation}
\frac{\delta z}{\delta t_0}=-\frac{1}{2}\Omega_{\rm m0}z+\mathcal O(z^2).
\end{equation}
The above equation shows that the redshift drift 
is non-positive near the center. 
This is the same behavior 
as that of the homogeneous and isotropic universe filled with dust.

\section{The redshift drift in LTB void models}
\label{sec:negative}

We call an LTB universe model the LTB {\it void} model, 
if the following three 
conditions are satisfied. 
\begin{enumerate}
\item the mass density is non-negative;  
\item the mass density is increasing with 
$r$ increasing in the domain $r>0$ on a 
spacelike hypersurface of constant $t$; 
\item $\delr R$ is positive;
\item $z$-normality. 
\end{enumerate}

\vskip0.5cm
\noindent
{\em Proposition} 1 \hskip0.2cmIn LTB void models, 
$\delt^2\delr R$ is negative. 

\vskip0.5cm
{\em Proof.} 
By Eq.~\eqref{eq:Einstein-eq}, we obtain 
\begin{eqnarray}
\delt^2\delr R(t,r)&=&-\frac{G\delr M}{R^2}
+\frac{2GM\delr R}{R^3} \nonumber \\
&=&4\pi G\frac{\delr R}{R^3}
\left(-\rho R^3+2\int_0^r\rho(t,x) R^2(t,x)\delr R(t,x) dx\right),
\end{eqnarray}
where we have used Eq.~\eqref{eq:rho} in the second equality. 
Since $\delr R$ is positive by the definition of LTB void models, 
we may replace the integration variable $x$ by $R=R(t,x)$ and obtain  
\begin{eqnarray}
\delt^2\delr R(t,r)
&=&4\pi G\frac{\delr R}{R^3}
\left(-\rho R^3+2\int_0^{R(t,r)}\rho R^2 dR\right) \nonumber \\
&=&-4\pi G\frac{\delr R}{R^3}
\int_0^{R(t,r)}\left(\frac{d\rho}{dR}R^3+\rho R^2\right)dR. 
\label{eq:forproof1}
\end{eqnarray}
Since $d\rho/dR=(\delr R)^{-1}\delr\rho$ is positive in the domain of $R>0$, 
the integrand in the last equality of the above equation is positive. \hfill Q.E.D.

\vskip0.5cm
\noindent
{\em Theorem} \hskip0.2cm In LTB void models, the redshift drift of 
an off-center source observed at the symmetry center is negative.

\vskip0.5cm
{\em Proof.} 
Since the cosmological redshift $z$ vanishes at $r=0$, 
$z$ is non-negative by the assumption of $z$-normality. 
Further, the $z$-normality leads to $\delt\delr R>0$ through Eq.~\eqref{eq:dzdr}. 
Then, since $\delta t_0>0$, we see from Eq.~\eqref{eq:delzfin} that 
Proposition 1 leads to the following inequality   
\begin{equation}
\frac{d}{dz}\left(\frac{\delta z}{1+z}\right) < 0.
\label{eq:forproof2}
\end{equation}
Since $\delta z$ should vanish at $z=0$, we have $\delta z<0$ 
for $z>0$ from the above 
inequality. \hfill Q.E.D.

\section{Redshift drift in LTB universe models with a large hump}
\label{sec:positive}

\begin{figure}[htbp]
\begin{center}
\includegraphics[scale=1.0]{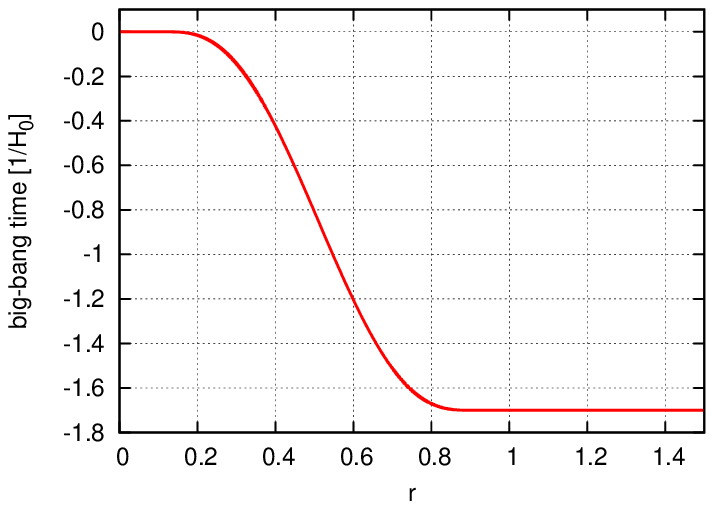}
\includegraphics[scale=1.0]{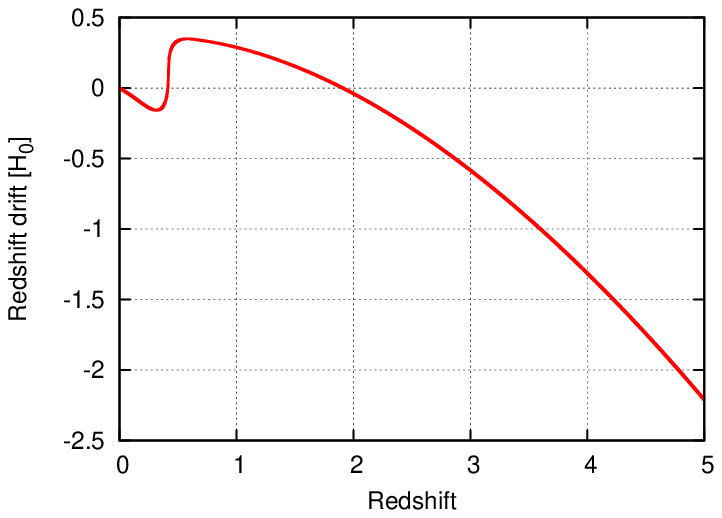}
\caption{
The LTB universe model with $k(r)=0$. 
The right panel depicts the redshift drift 
$\delta z/\delta t_0$ of the LTB universe model with $k(r)=0$ 
as a function of the redshift $z$. 
The other arbitrary function $t_{\rm B}(r)$ is shown in the left panel. 
}
\label{fig:dzdecay}
\end{center}
\end{figure}

\begin{figure}[htbp]
\begin{center}
\includegraphics[scale=1.0]{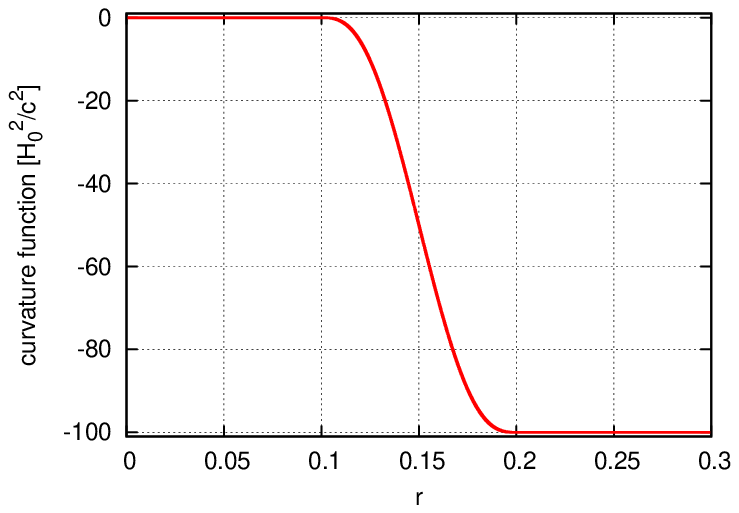}
\includegraphics[scale=1.0]{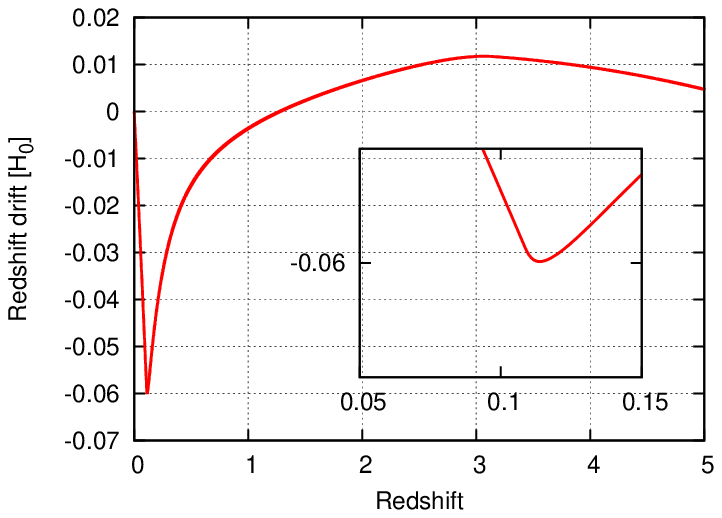}
\caption{
The right panel depicts the redshift drift 
$\delta z/\delta t_0$ of the LTB universe model with $t_{\rm B}(r)=0$ 
as a function of the redshift $z$. 
The other arbitrary function $k(r)$ is shown in the left panel. 
}
\label{fig:dzgrow}
\end{center}
\end{figure}

In the preceding section, we showed that the redshift drift observed at the 
symmetry center is negative for $r>0$ in LTB void models. 
Conversely, if there is a domain in which the 
mass density is decreasing with increasing $r$, the 
redshift drift might be negative. 
In this section, we show that it is true with hump-type 
mass density distributions. 
We consider the following two LTB universe models,
\begin{itemize}
\item[(i)]{$k(r)=0$ and $\tb(r)=f(r;a,r_1,r_2)$\\ with $a=-1.7H_0^{-1}$, 
$r_1=0.12cH_0^{-1}$ and $r_2=0.9cH_0^{-1}$,}
\item[(ii)]{
$\tb(r)=0$ and $k(r)=f(r;a,r_1,r_2)$\\ with $a=-100c^{-2}H_0^2$, 
$r_1=0.1cH_0^{-1}$ and $r_2=0.2cH_0^{-1}$, }
\end{itemize}
where 
\begin{equation}
f(r;a,r_1,r_2)=
\left\{
\begin{array}{lll}
0~~&{\rm for}~~r<r_1\,,\\
\displaystyle
\frac{a\left(r-r_1\right)^3
\left(r_1^2-5r_1r_2+10r_2^2+3r_1r-15r_2r+6r^2\right)}{(r_2-r_1)^5}
~~&{\rm for}~~r_1\leq  r<r_2\,,\\
a~~&{\rm for}~~r_2\leq r\,.
\end{array}
\right.
\end{equation}
In Figs.\ref{fig:dzdecay} and \ref{fig:dzgrow}, 
we show the redshift drifts 
of these models. 
Although we do not show the energy densities of these models, 
a large hump in the mass density distribution exists in each model as well as 
in $t_{\rm B}(r)$ or $k(r)$. 
Although there is a redshift domain with 
positive redshift drift in each example, the distance-redshift 
relations of these models do not agree with the observational data, and 
further, the inhomogeneities need to be very large. 


\section{summary and discussion}
\label{sec:conclusions}

In this paper, we studied the redshift drift in LTB universe models 
in which the observer is located at the symmetry center. 
We showed that, assuming that the mass density of the dust is positive, 
the redshift drift of an off-center source 
is negative if the mass density and the circumferential radius are 
increasing functions of the comoving radial coordinate. 
We also showed that if there is a very large hump structure around the 
symmetry center, the redshift drift can be positive. 
As a result, by observation of the redshift drift, 
we get a strong constraint 
on void-type universe models: 
if the redshift drift turns out 
to be positive in some redshift domain, 
LTB void models can be rejected.

By projects such as 
the Cosmic Dynamics 
Experiment(CODEX)~\cite{2005Msngr.122...10P,Cristiani:2007by,Liske:2008ph} 
in the European Extremely Large Telescope(E-ELT)\cite{2007Msngr.127...11G} and 
PREcision Super Stable Observations (ESPRESSO) in the Very Large Telescope 
array(VLT)~\cite{Cristiani:2007by,Liske:2008ph,D'Odorico:2007qr}, 
it is possible to get highly accurate spectroscopic observational data. 
The observability of the redshift drift by CODEX has been analyzed in 
Refs.~\cite{Corasaniti:2007bg,Balbi:2007fx,Cristiani:2007by,Liske:2008ph}, and 
the test of LTB universe models by this project has been investigated in 
Ref.~\cite{Quartin:2009xr}. 
Targets of these projects are QSOs whose redshifts are larger than two. 
On the other hand, observing the sign of the redshift drift in the 
redshift domain $z\lesssim 2$ is very crucial to test LTB void universe models 
in contrast with Copernican universe models. 
This is because known examples of Copernican universe models with dark energy or 
modified gravity predict a positive redshift drift 
in the redshift domain $z\lesssim 2$, 
while it is negative for LTB void models. 
From, for example, the observation of compact binary stars by 
DECIGO~\cite{Seto:2001qf,Kawamura:2006up,Sato:2009zzb} or 
BBO~\cite{Ungarelli:2005qb,BBO:2003}, the redshift drift at $z \simeq 1$ can be 
measured~\cite{Seto:2001qf}. 
Observations by DECIGO or BBO over several years will make it possible to 
observe the sign of the redshift drift for $z\lesssim 2$ and to
test LTB void models~\cite{zdriftdecigo}.

\section*{Acknowledgements}
We thank A. Nishizawa and K. Yagi for helpful discussions and comments.
We would also like to thank all the participants of the 
Long-term Workshop on Gravity and Cosmology (GC2010: YITP-T-10-01)
for fruitful discussions. 
This work is supported in part by JSPS Grant-in-Aid for  Creative
Scientific Research No.~19GS0219 and JSPS Grant-in-Aid for Scientific
Research (C)  No.~21540276.


\begin{thebibliography}{10}

\bibitem{Zehavi:1998gz}
I.~Zehavi, A.~G. Riess, R.~P. Kirshner, and A.~Dekel,
\newblock Astrophys. J. {\bf 503}, 483 (1998), arXiv:astro-ph/9802252, {\em {A
  Local Hubble Bubble from SNe Ia?}}

\bibitem{Tomita:1999qn}
K.~Tomita,
\newblock Astrophys. J. {\bf 529}, 38 (2000), arXiv:astro-ph/9906027, {\em
  {Distances and lensing in cosmological void models}}.

\bibitem{Tomita:2000jj}
K.~Tomita,
\newblock Mon. Not. Roy. Astron. Soc. {\bf 326}, 287 (2001),
  arXiv:astro-ph/0011484, {\em {A Local Void and the Accelerating Universe}}.

\bibitem{Tomita:2001gh}
K.~Tomita,
\newblock Prog. Theor. Phys. {\bf 106}, 929 (2001), arXiv:astro-ph/0104141,
  {\em {Analyses of Type Ia Supernova Data in Cosmological Models with a Local
  Void}}.

\bibitem{Celerier:1999hp}
M.-N. Celerier,
\newblock Astron. Astrophys. {\bf 353}, 63 (2000), arXiv:astro-ph/9907206, {\em
  {Do we really see a cosmological constant in the supernovae data ?}}

\bibitem{Alnes:2005rw}
H.~Alnes, M.~Amarzguioui, and O.~Gron,
\newblock Phys. Rev. {\bf D73}, 083519 (2006), arXiv:astro-ph/0512006, {\em {An
  inhomogeneous alternative to dark energy?}}

\bibitem{Alnes:2006pf}
H.~Alnes and M.~Amarzguioui,
\newblock Phys. Rev. {\bf D74}, 103520 (2006), arXiv:astro-ph/0607334, {\em
  {CMB anisotropies seen by an off-center observer in a spherically symmetric
  inhomogeneous universe}}.

\bibitem{Alnes:2006uk}
H.~Alnes and M.~Amarzguioui,
\newblock Phys. Rev. {\bf D75}, 023506 (2007), arXiv:astro-ph/0610331, {\em
  {The supernova Hubble diagram for off-center observers in a spherically
  symmetric inhomogeneous universe}}.

\bibitem{Alexander:2007xx}
S.~Alexander, T.~Biswas, A.~Notari, and D.~Vaid,
\newblock JCAP {\bf 0909}, 025 (2009), arXiv:0712.0370, {\em {Local Void vs
  Dark Energy: Confrontation with WMAP and Type Ia Supernovae}}.

\bibitem{Bolejko:2008cm}
K.~Bolejko and J.~S.~B. Wyithe,
\newblock JCAP {\bf 0902}, 020 (2009), arXiv:0807.2891, {\em {Testing the
  Copernican Principle via Cosmological Observations}}.

\bibitem{Enqvist:2007vb}
K.~Enqvist,
\newblock Gen. Rel. Grav. {\bf 40}, 451 (2008), arXiv:0709.2044, {\em
  {Lemaitre-Tolman-Bondi model and accelerating expansion}}.

\bibitem{Enqvist:2006cg}
K.~Enqvist and T.~Mattsson,
\newblock JCAP {\bf 0702}, 019 (2007), arXiv:astro-ph/0609120, {\em {The effect
  of inhomogeneous expansion on the supernova observations}}.

\bibitem{February:2009pv}
S.~February, J.~Larena, M.~Smith, and C.~Clarkson,
\newblock (2009), arXiv:0909.1479, {\em {Rendering Dark Energy Void}}.

\bibitem{GarciaBellido:2008nz}
J.~Garcia-Bellido and T.~Haugboelle,
\newblock JCAP {\bf 0804}, 003 (2008), arXiv:0802.1523, {\em {Confronting
  Lemaitre-Tolman-Bondi models with Observational Cosmology}}.

\bibitem{GarciaBellido:2008yq}
J.~Garcia-Bellido and T.~Haugboelle,
\newblock JCAP {\bf 0909}, 028 (2009), arXiv:0810.4939, {\em {The radial BAO
  scale and Cosmic Shear, a new observable for Inhomogeneous Cosmologies}}.

\bibitem{GarciaBellido:2008gd}
J.~Garcia-Bellido and T.~Haugboelle,
\newblock JCAP {\bf 0809}, 016 (2008), arXiv:0807.1326, {\em {Looking the void
  in the eyes - the kSZ effect in LTB models}}.

\bibitem{Garfinkle:2006sb}
D.~Garfinkle,
\newblock Class. Quant. Grav. {\bf 23}, 4811 (2006), arXiv:gr-qc/0605088, {\em
  {Inhomogeneous spacetimes as a dark energy model}}.

\bibitem{Kasai:2007fn}
M.~Kasai,
\newblock Prog. Theor. Phys. {\bf 117}, 1067 (2007), arXiv:astro-ph/0703298,
  {\em {Apparent Acceleration through Large-scale Inhomogeneities
  --Post-Friedmannian Effects of Inhomogeneities on the Luminosity
  Distance--}}.

\bibitem{2000ApJ...529...26T}
K.~{Tomita},
\newblock Astrophys. J. {\bf 529}, 26 (2000), arXiv:astro-ph/9905278, {\em
  {Bulk Flows and Cosmic Microwave Background Dipole Anisotropy in Cosmological
  Void Models}}.

\bibitem{Tomita:2000rf}
K.~Tomita,
\newblock Prog. Theor. Phys. {\bf 105}, 419 (2001), arXiv:astro-ph/0005031,
  {\em {Anisotropy of the Hubble Constant in a Cosmological Model with a Local
  Void on Scales of ~ 200 Mpc}}.

\bibitem{Zibin:2008vk}
J.~P. Zibin, A.~Moss, and D.~Scott,
\newblock Phys. Rev. Lett. {\bf 101}, 251303 (2008), arXiv:0809.3761, {\em {Can
  we avoid dark energy?}}

\bibitem{Clifton:2009kx}
T.~Clifton, P.~G. Ferreira, and J.~Zuntz,
\newblock JCAP {\bf 0907}, 029 (2009), arXiv:0902.1313, {\em {What the small
  angle CMB really tells us about the curvature of the Universe}}.

\bibitem{Biswas:2007gi}
T.~Biswas and A.~Notari,
\newblock JCAP {\bf 0806}, 021 (2008), arXiv:astro-ph/0702555, {\em
  {Swiss-Cheese Inhomogeneous Cosmology \& the Dark Energy Problem}}.

\bibitem{Bolejko:2005fp}
K.~Bolejko,
\newblock PMC Phys. {\bf A2}, 1 (2008), arXiv:astro-ph/0512103, {\em
  {Supernovae Ia observations in the Lemaitre--Tolman model}}.

\bibitem{Bolejko:2008xh}
K.~Bolejko,
\newblock Gen. Rel. Grav. {\bf 41}, 1737 (2009), arXiv:0804.1846, {\em {The
  Szekeres Swiss Cheese model and the CMB observations}}.

\bibitem{Bolejko:2008ya}
K.~Bolejko and P.~Lasky,
\newblock (2008), arXiv:0809.0334, {\em {Pressure gradients, shell crossing
  singularities and acoustic oscillations - application to inhomogeneous
  cosmological models}}.

\bibitem{Brouzakis:2006dj}
N.~Brouzakis, N.~Tetradis, and E.~Tzavara,
\newblock JCAP {\bf 0702}, 013 (2007), arXiv:astro-ph/0612179, {\em {The Effect
  of Large-Scale Inhomogeneities on the Luminosity Distance}}.

\bibitem{Brouzakis:2007zi}
N.~Brouzakis, N.~Tetradis, and E.~Tzavara,
\newblock JCAP {\bf 0804}, 008 (2008), arXiv:astro-ph/0703586, {\em {Light
  Propagation and Large-Scale Inhomogeneities}}.

\bibitem{Bhattacharya:2009bz}
S.~Bhattacharya, P.~S. Joshi, and K.-i. Nakao,
\newblock (2009), arXiv:0911.2297, {\em {Accelerated cosmic expansion in a
  scalar-field universe}}.

\bibitem{Clarkson:2007pz}
C.~Clarkson, B.~Bassett, and T.~H.-C. Lu,
\newblock Phys. Rev. Lett. {\bf 101}, 011301 (2008), arXiv:0712.3457, {\em {A
  general test of the Copernican Principle}}.

\bibitem{Caldwell:2007yu}
R.~R. Caldwell and A.~Stebbins,
\newblock Phys. Rev. Lett. {\bf 100}, 191302 (2008), arXiv:0711.3459, {\em {A
  Test of the Copernican Principle}}.

\bibitem{Clifton:2008hv}
T.~Clifton, P.~G. Ferreira, and K.~Land,
\newblock Phys. Rev. Lett. {\bf 101}, 131302 (2008), arXiv:0807.1443, {\em
  {Living in a Void: Testing the Copernican Principle with Distant
  Supernovae}}.

\bibitem{Dabrowski:1997sm}
M.~P. Dabrowski and M.~A. Hendry,
\newblock Astrophys. J. {\bf 498}, 67 (1998), arXiv:astro-ph/9704123, {\em
  {Non-Uniform Pressure Universes: The Hubble Diagram of Type Ia Supernovae and
  the Age of the Universe}}.

\bibitem{Goodman:1995dt}
J.~Goodman,
\newblock Phys. Rev. {\bf D52}, 1821 (1995), arXiv:astro-ph/9506068, {\em
  {Geocentrism reexamined}}.

\bibitem{Godlowski:2004gh}
W.~Godlowski, J.~Stelmach, and M.~Szydlowski,
\newblock Class. Quant. Grav. {\bf 21}, 3953 (2004), arXiv:astro-ph/0403534,
  {\em {Can the Stephani model be an alternative to FRW accelerating models?}}

\bibitem{Jia:2008ti}
J.~Jia and H.-b. Zhang,
\newblock JCAP {\bf 0812}, 002 (2008), arXiv:0809.2597, {\em {Can the
  Copernican principle be tested by cosmic neutrino background?}}

\bibitem{Lasky:2010vn}
P.~D. Lasky and K.~Bolejko,
\newblock (2010), arXiv:1001.1159, {\em {The effect of pressure gradients on
  luminosity distance - redshift relations}}.

\bibitem{Marra:2007pm}
V.~Marra, E.~W. Kolb, S.~Matarrese, and A.~Riotto,
\newblock Phys. Rev. {\bf D76}, 123004 (2007), arXiv:0708.3622, {\em {On
  cosmological observables in a swiss-cheese universe}}.

\bibitem{Moffat:2006ct}
J.~W. Moffat,
\newblock (2006), arXiv:astro-ph/0606124, {\em {Inhomogeneous Cosmology,
  Inflation and Late-Time Accelerating Universe}}.

\bibitem{PascualSanchez:1999zr}
J.~F. Pascual-Sanchez,
\newblock Mod. Phys. Lett. {\bf A14}, 1539 (1999), arXiv:gr-qc/9905063, {\em
  {Cosmic acceleration: Inhomogeneity versus vacuum energy}}.

\bibitem{Quartin:2009xr}
M.~Quartin and L.~Amendola,
\newblock (2009), arXiv:0909.4954, {\em {Distinguishing Between Void Models and
  Dark Energy with Cosmic Parallax and Redshift Drift}}.

\bibitem{Romano:2007zz}
A.~E. Romano,
\newblock Phys. Rev. {\bf D76}, 103525 (2007), {\em {Redshift spherical shell
  energy in isotropic universes}}.

\bibitem{Stelmach:2006zc}
J.~Stelmach and I.~Jakacka,
\newblock Class. Quant. Grav. {\bf 23}, 6621 (2006), {\em {Angular sizes in
  spherically symmetric Stephani cosmological models}}.

\bibitem{Tanimoto:2009mz}
M.~Tanimoto, Y.~Nambu, and K.~Iwata,
\newblock (2009), arXiv:0906.4857, {\em {The Role of Anisotropy in the Void
  Models without Dark Energy}}.

\bibitem{Tomita:2009yx}
K.~Tomita,
\newblock (2009), arXiv:0912.4773, {\em {Gauge-invariant treatment of the
  integrated Sachs-Wolfe effect on general spherically symmetric spacetimes}}.

\bibitem{Uzan:2008qp}
J.-P. Uzan, C.~Clarkson, and G.~F.~R. Ellis,
\newblock Phys. Rev. Lett. {\bf 100}, 191303 (2008), arXiv:0801.0068, {\em
  {Time drift of cosmological redshifts as a test of the Copernican
  principle}}.

\bibitem{Yoo:2010qy}
C.-M. Yoo, K.-i. Nakao, and M.~Sasaki,
\newblock (2010), arXiv:1005.0048, {\em {CMB observations in LTB universes:
  Part I: Matching peak positions in the CMB spectrum}}.

\bibitem{Clarkson:1999yj}
C.~A. Clarkson and R.~Barrett,
\newblock Class. Quant. Grav. {\bf 16}, 3781 (1999), arXiv:gr-qc/9906097, {\em
  {Does the Isotropy of the CMB Imply a Homogeneous Universe? Some Generalised
  EGS Theorems}}.

\bibitem{Zibin:2008vj}
J.~P. Zibin,
\newblock ~  (2008), arXiv:0804.1787, {\em {Scalar Perturbations on
  Lemaitre-Tolman-Bondi Spacetimes}}.

\bibitem{Regis:2010iq}
M.~Regis and C.~Clarkson,
\newblock (2010), arXiv:1003.1043, {\em {Do primordial Lithium abundances imply
  there's no Dark Energy?}}

\bibitem{Clarkson:2010ej}
C.~Clarkson and M.~Regis,
\newblock (2010), arXiv:1007.3443, {\em {The Cosmic Microwave Background in an
  Inhomogeneous Universe}}.

\bibitem{Clarkson:2010uz}
C.~Clarkson and R.~Maartens,
\newblock Class. Quant. Grav. {\bf 27}, 124008 (2010), arXiv:1005.2165, {\em
  {Inhomogeneity and the foundations of concordance cosmology}}.

\bibitem{Biswas:2010xm}
T.~Biswas, A.~Notari, and W.~Valkenburg,
\newblock (2010), arXiv:1007.3065, {\em {Testing the Void against Cosmological
  data: fitting CMB, BAO, SN and H0}}.

\bibitem{Kodama:2010gr}
H.~Kodama, K.~Saito, and A.~Ishibashi,
\newblock (2010), arXiv:1004.3089, {\em {Analytic formulae for the off-center
  CMB anisotropy in a general spherically symmetric universe}}.

\bibitem{Foreman:2010uj}
S.~Foreman, A.~Moss, J.~P. Zibin, and D.~Scott,
\newblock (2010), arXiv:1009.0273, {\em {Spatial and temporal tuning in void
  models for acceleration}}.

\bibitem{Moss:2010jx}
A.~Moss, J.~P. Zibin, and D.~Scott,
\newblock (2010), arXiv:1007.3725, {\em {Precision Cosmology Defeats Void
  Models for Acceleration}}.

\bibitem{Marra:2010pg}
V.~Marra and M.~Paakkonen,
\newblock (2010), arXiv:1009.4193, {\em {Observational constraints on the LLTB
  model}}.

\bibitem{Zhang:2010fa}
P.~Zhang and A.~Stebbins,
\newblock (2010), arXiv:1009.3967, {\em {Confirmation of the Copernican
  principle at Gpc radial scale and above from the kinetic Sunyaev Zel'dovich
  effect power spectrum}}.

\bibitem{Dunsby:2010ts}
P.~Dunsby, N.~Goheer, B.~Osano, and J.-P. Uzan,
\newblock (2010), arXiv:1002.2397, {\em {How close can an Inhomogeneous
  Universe mimic the Concordance Model?}}

\bibitem{Yoo:2008su}
C.-M. Yoo, T.~Kai, and K.-i. Nakao,
\newblock Prog. Theor. Phys. {\bf 120}, 937 (2008), arXiv:0807.0932, {\em
  {Solving Inverse Problem with Inhomogeneous Universe}}.

\bibitem{Misner:1964je}
C.~W. Misner and D.~H. Sharp,
\newblock Phys. Rev. {\bf 136}, B571 (1964), {\em {Relativistic equations for
  adiabatic, spherically symmetric gravitational collapse}}.

\bibitem{Tanimoto:2007dq}
M.~Tanimoto and Y.~Nambu,
\newblock Class. Quant. Grav. {\bf 24}, 3843 (2007), arXiv:gr-qc/0703012, {\em
  {Luminosity distance-redshift relation for the LTB solution near the
  center}}.

\bibitem{2005Msngr.122...10P}
L.~{Pasquini} {\em et~al.},
\newblock The Messenger {\bf 122}, 10 (2005), {\em {CODEX: Measuring the
  Expansion of the Universe (and beyond)}}.

\bibitem{Cristiani:2007by}
S.~Cristiani {\em et~al.},
\newblock Nuovo Cim. {\bf 122B}, 1159 (2007), arXiv:0712.4152, {\em {The
  CODEX-ESPRESSO experiment: cosmic dynamics, fundamental physics, planets and
  much more.}}

\bibitem{Liske:2008ph}
J.~Liske {\em et~al.},
\newblock Mon. Not. Roy. Astron. Soc. {\bf 386}, 1192 (2008), arXiv:0802.1532,
  {\em {Cosmic dynamics in the era of Extremely Large Telescopes}}.

\bibitem{2007Msngr.127...11G}
R.~{Gilmozzi} and J.~{Spyromilio},
\newblock The Messenger {\bf 127}, 11 (2007), {\em {The European Extremely
  Large Telescope (E-ELT)}}.

\bibitem{D'Odorico:2007qr}
CODEX/ESPRESSO, V.~D'Odorico,
\newblock (2007), arXiv:0708.1258, {\em {CODEX/ESPRESSO: the era of precision
  spectroscopy}}.

\bibitem{Corasaniti:2007bg}
P.-S. Corasaniti, D.~Huterer, and A.~Melchiorri,
\newblock Phys. Rev. {\bf D75}, 062001 (2007), arXiv:astro-ph/0701433, {\em
  {Exploring the Dark Energy Redshift Desert with the Sandage-Loeb Test}}.

\bibitem{Balbi:2007fx}
A.~Balbi and C.~Quercellini,
\newblock (2007), arXiv:0704.2350, {\em {The time evolution of cosmological
  redshift as a test of dark energy}}.

\bibitem{Seto:2001qf}
N.~Seto, S.~Kawamura, and T.~Nakamura,
\newblock Phys. Rev. Lett. {\bf 87}, 221103 (2001), arXiv:astro-ph/0108011,
  {\em {Possibility of direct measurement of the acceleration of the universe
  using 0.1-Hz band laser interferometer gravitational wave antenna in space}}.

\bibitem{Kawamura:2006up}
S.~Kawamura {\em et~al.},
\newblock Class. Quant. Grav. {\bf 23}, S125 (2006), {\em {The Japanese space
  gravitational wave antenna DECIGO}}.

\bibitem{Sato:2009zzb}
S.~Sato {\em et~al.},
\newblock J. Phys. Conf. Ser. {\bf 154}, 012040 (2009), {\em {DECIGO: The
  Japanese space gravitational wave antenna}}.

\bibitem{Ungarelli:2005qb}
C.~Ungarelli, P.~Corasaniti, R.~A. Mercer, and A.~Vecchio,
\newblock Class. Quant. Grav. {\bf 22}, S955 (2005), arXiv:astro-ph/0504294,
  {\em {Gravitational waves, inflation and the cosmic microwave background:
  towards testing the slow-roll paradigm}}.

\bibitem{BBO:2003}
E.~S. Phinney {\em et~al.},
\newblock (NASA)  (2003), {\em {Big Bang Observer Mission Concept Study}}.

\bibitem{zdriftdecigo}
K.~Yagi, A.~Nishizawa, and C.~Yoo,
\newblock in preparation .

\end{thebibliography}
%
\end{document}